\documentclass[conference]{IEEEtran}
\ifCLASSINFOpdf
\else
\fi
\usepackage[utf8]{inputenc}
\usepackage[english]{babel}

\usepackage[linesnumbered,ruled]{algorithm2e}

\usepackage[acronym]{glossaries}

\usepackage{bbm}
\usepackage{amsmath}
\usepackage{mathtools}
\usepackage{amssymb}
\usepackage{amsthm}
\usepackage{comment}
\usepackage{pifont}

\DeclareMathOperator*{\argmin}{arg\,min}
\usepackage{multirow}
\usepackage{graphicx}
\usepackage{epsfig}
\usepackage{color}
\usepackage{dsfont}
\usepackage{soul}
\usepackage{hyperref}

\DeclareMathAlphabet\mathbfcal{OMS}{cmsy}{b}{n}

\def\trace{\mathrm{tr}}

\def\Sigmabar{\bar{\Sigma}}
\def\Sigmahat{\widehat{\Sigma}}
\def\Sigtilde{\widetilde{\Sigma}}
\def\Sigbold{\boldsymbol{\Sigma}}

\def\muhat{\hat{\mu}}

\def\mubold{\boldsymbol{\mu}}

\def\wbar{\bar{w}}
\def\wtilde{\tilde{w}}
\def\mubar{\bar{\mu}}
\def\mutilde{\tilde{\mu}}

\def\ones{ \mathds{1} }
\def\wbold{\boldsymbol{w}}

\def\KL{\textit{KL}}
\def\FKL{\textit{FKL}}
\def\RKL{\textit{RKL}}
\def\SKL{\textit{SKL}}

\def\CS{\textit{CS}}

\def\DKL{D_\KL}
\def\DFKL{D_\FKL}
\def\DRKL{D_\RKL}
\def\DSKL{D_\SKL}

\def\DL2{D_{L\!2}}
\def\DCS{D_\CS}
\def\DJR2{D_{J\!R\!D}^2}

\def\DB{D_B}

\def\DW2{D_{W\!2}}
\def\DH2{D_{H\!2}}
\def\DP{D_{P}}
\def\DN{D_{N}}
\def\DJalpha{D_{J}^{\alpha}}
\def\DJ{D_{J}}

\def\DaP{D^\prime_{\alpha}}
\def\DaS{D^{\prime \prime}_{\alpha}}

\def\DLB{D_{L\!B}}

\def\Real{\mathbb{R}}

\def\wtilde{\tilde{w}}

\def\nuhat{\hat{\nu}}

\def\qhat{\hat{q}}

\def\Qcal{\mathcal{Q}}
\def\nubold{\boldsymbol{\nu}}

\def\qbold{\boldsymbol{q}}

\def\qbold{\boldsymbol{q}}

\def\mutildeKLA{\mutilde_\textit{KLA}}
\def\PsiMPM{\Psi_\textit{MP}}
\def\muMPM{\mu_\textit{MP}}
\def\PsiKLA{\Psi_\textit{KLA}}
\def\muKLA{\mu_\textit{KLA}}
\def\GammaKLA{\Gamma_\textit{KLA}}
\def\GammaMPM{\Gamma_\textit{MP}}

\def\d{\mathrm{d}}

\newtheorem{rem}{Remark}

\newcommand\blfootnote[1]{%
  \begingroup
  \renewcommand\thefootnote{}\footnote{#1}%
  \addtocounter{footnote}{-1}%
  \endgroup
}

\usepackage[noadjust]{cite}

\begin{document}
%
\title{ Fixed-point iterative computation of Gaussian barycenters for some dissimilarity measures \\
\begin{normalsize}
    Regular Research Paper - CSCI-RTPC Research Track
\end{normalsize}}

\author{\IEEEauthorblockN{Alessandro D'Ortenzio$^{1,2}$\ \ , \ \ Costanzo Manes$^{1,2}$\ \ and \ \ Umut Orguner$^{3}$}
\IEEEauthorblockA{ 
$^1$Department of Information Engineering, Computer Science and Mathematics, University of L'Aquila\\
$^2$Center of Excellence EX-EMERGE, University of L'Aquila\\
$^3$Department of Electrical and Electronics Engineering, Middle East Technical University, Turkey\\
Email: alessandro.dortenzio@univaq.it$^{1,2}$,  \ costanzo.manes@univaq.it$^{1,2}$, \ 
umut@metu.edu.tr$^3$}
}

\maketitle

\def\hide#1{}

\begin{abstract}
In practical contexts like sensor fusion or computer vision, it is not unusual to deal with a large number of Gaussian densities that encode the available information.
In such cases, if the computational capabilities are limited, a data compression is required, often done by finding the barycenter of the set of Gaussians.
However, such computation strongly depends on the chosen loss function (dissimilarity measure) to be minimized, and most often it must be performed by means of numerical methods, since the barycenter can rarely be computed analytically.
Some constraints, like the covariance matrix symmetry and positive definiteness can make non-trivial the numerical computation of the Gaussian barycenter. 
In this work, a set of Fixed-Point Iteration algorithms are presented in order to allow for the agile computation of Gaussian barycenters according to several dissimilarity measures.
\end{abstract}
\begin{IEEEkeywords}
Gaussian densities, barycenters, fixed-point iterations, data compression, signal processing.
\end{IEEEkeywords}

\blfootnote{This work was supported by the Centre of EXcellence on Connected, Geo-Localized and Cybersecure Vehicles (EX-EMERGE), funded by the Italian Government under CIPE resolution n. 70/2017 (Aug.\ 7, 2017).}

\IEEEpeerreviewmaketitle
\vspace{-0.6cm}
\section{Introduction}
The Gaussian density is a commonly used parametric model in many applications, mainly because it can be a reasonable approximation for many random phenomena.
In general, such a model allows for a compact and efficient way to represent uncertainty, hence being suitable even for real-time applications. 
Nonetheless, when dealing with complex phenomena, it may be necessary to consider sets of Gaussians in order to properly describe the corresponding uncertainty, which may be computationally intractable or inefficient for many elaboration systems. 
Some examples of usage of sets of Gaussian densities are: 
target tracking in the presence of clutter and filtering of stochastic switching systems, 
where the Gaussians represents different \textit{hypotheses}, 
fusion of information from several sensors, image retrieval, data clustering (\cite{MMA2000,RobustEM}) and so on. 
In those problems, the idea of finding a single representative for a set of Gaussian densities is pursued in order to reduce the representation complexity, while preserving most of the available information. 
One approach to achieve this is by computing the barycenter of the involved components, obtained by minimizing the average dissimilarity from each hypothesis in the set; nonetheless, except for very few cases where closed forms are available, this may be a non-trivial task \cite{ClaiciICML18,BurbeaRao,BregCentroids}.
The outcome of such a problem strongly depends on the chosen loss function, that in general is a dissimilarity measure between probability density functions (pdfs), and in most cases it is necessary to resort to numerical optimization. 
In addition, due to reasons like the presence of constraints on the covariance matrix (symmetry and positive definiteness (SPD)), standard optimization methods, e.g., the gradient descent, may not be well suited to address this task. A viable alternative is to consider Fixed-Point Iteration (FPI) methods which often provide compact and efficient recursions, and which can be very effective in the barycenter evaluation.

\noindent 
The main goal of this work is to provide and discuss an ensemble of Fixed-Point Iteration (FPI) algorithms for several dissimilarity measures that allow for the agile computation of barycenters in the Gaussian case.
The motivation behind this work relies on the necessity to find more compact and efficient ways to evaluate barycenters when sets/mixtures of Gaussians are considered. In the consistent Mixture Reduction framework proposed in \cite{CGMR,Consistency}, the main bottleneck can be represented by the computational load introduced in the barycenter evaluation; for this reason, exploring efficient solutions to the barycenter problem may improve the applicability of such concepts. 
Nonetheless, the topic of mixture reduction is beyond the scope of this paper, so the focus will be limited to discuss the barycenter computation for sets of Gaussians from a general perspective.

\noindent The paper is organized as follows:
in Sec.~\ref{sec:barycenters} the barycenter of a weighted set of distributions is formally defined and its computation problem is formulated; 
in Sec.~\ref{sec:baryGaussians} the problem is studied for several dissimilarity measures in the Gaussian case, providing in parallel the corresponding fixed-point iterations. 
In Sec.~\ref{sec:discussion} a discussion regarding initialization, convergence and properties of the proposed algorithms is reported. Conclusions follow.

\noindent {\it Notations:}
in this work, $\Real^n_+$ denotes the set of nonnegative vectors in $\Real^n$, $I_n$  is the identity matrix in $\Real^{n\times n}$ while $\ones_n$ denotes a vector of ones in $\Real^n$.
$S^d\subset\Real^{d\times d}$ denotes the open cone of symmetric positive definite $d\times d$ matrices.

\section{Barycenters of weighted sets of distributions} \label{sec:barycenters}

The notion of barycenter emerges when it is required to find a single distribution that is {\it close}, in some sense, to a set of weighted distributions.
A case of particular interest is the one where all distributions are in a given class, and the barycenter is sought in the same class \cite{ClaiciICML18,BurbeaRao,BregCentroids}. 
In applications where the number of densities becomes large, finding a representative which encodes the available information with minimal loss plays a key role. 
In this regard, consider a set of probability density functions  (pdfs) $q_i(x)$, $x\in\Real^d$, $i=1,\dots, n$, and a set of positive weights $w_i$, each one associated to a distribution in the set.
The set of pairs $(w_i,q_i)$, $i=1,...n$, is denoted as $(\wbold,\qbold)$.
In particular, we will consider pdfs $q_i(x)$ in a given class $\Qcal$ of distributions. 

In order to formulate the barycenter problem, the concept of \textit{dissimilarity}, or loss, is required.
There exist quite a lot of dissimilarity/loss measures ($D$-measure for short) between distributions in the literature, often also called {\it distorsion} measures or {\it statistical divergences}; some of the most used $D$-measures are considered in this paper, and are listed in the Appendix.
The symbol $D(p \Vert q)$ denotes a generic $D$-measure between 
two distributions $p$ and $q$ on the same probability space.
In general, $D$-measures must satisfy the two following requirements, for all pairs of distributions $p$ and $q$:

\vspace{-0.3cm}
\begin{small}
\begin{align}
& D(p\Vert q)\ge 0,\quad & & \text{\small (nonnegativity)}; \label{eq:Dnonneg}\\  
& D(p\Vert q) = 0\ \Longleftrightarrow\ p=q, & & \text{\small (identity of indiscernibles)}.\label{eq:Dident}
\end{align}
\end{small}
\vspace{-0.4cm}

If a given $D$-measure is not symmetric, that is $D(p\Vert q) \neq D(q\Vert p)$, it makes sense to define the {\it reverse} $D$-measure as $D^R(p\Vert q)=D(q\Vert p)$. 

Given a weighted set of distributions $(\wbold,\qbold)$ and a distribution 
$q$, an {\it Average Dissimilarity Function} (ADF) according to some chosen $D$-measure, can be defined as follows:

\vspace{-0.3cm}
\begin{small}
\begin{equation}   \label{eq:mDofp}
    m_D(q|\wbold,\qbold) = \frac{1}{\wbold^T \ones_n}\sum_{i=1}^n w_i D(q_i\Vert q).
\end{equation}
\end{small}
\vspace{-0.3cm}

\noindent The barycenter $\qhat$ of the weighted set $(\wbold,\qbold)$ is defined as the distribution
that minimizes the ADF \eqref{eq:mDofp}.
Mostly, it is of interest to find the barycenter in the same class $\Qcal$ of the distributions $q_i$, so that

\vspace{-0.3cm}
\begin{small}
\begin{equation} \label{eq:barycenterdefGEN}
\qhat= \argmin_{q\in\Qcal} m_D(q|\wbold,\qbold).
\end{equation}
\end{small}
\vspace{-0.5cm}

$\qhat$ is the distribution that, in average, is the {\it most similar} to all the distributions in the set $(\wbold,\qbold)$.
When the weights $w_i$ are all equal, the barycenter is also called the {\it centroid} of the set
\cite{BurbeaRao, BregCentroids}.
It is important to note that the barycenter of a set $(\wbold,\qbold)$
strongly depends on the choice of the $D$-measure, as it will be discussed
throughout this paper.

\begin{rem} \label{rem:barycenterRev}
If the considered $D$-measure lacks symmetry, a different ADF can be defined by reversing the order of the distributions $q_i$ and $q$ in \eqref{eq:mDofp}. The corresponding \textit{reverse} ADF will be denoted as $m^R_D$.
In the general case of non symmetric $D$-measures, 
$m_D$ and $m_D^R$ do not coincide; accordingly, the forward and reverse barycenters can be very different.
\end{rem}


\section{Barycenters of Weighted Sets of Gaussians}\label{sec:baryGaussians}

In this work the symbol $\nu(x\vert \mu, \Sigma)$ is used to denote a multivariate non-degenerate $d$-dimensional Gaussian density with mean $\mu\in\Real^d$ and covariance $\Sigma\in S^d$, of the form:

\vspace{-0.15cm}
\begin{small}
\begin{equation}\label{eq:GaussianDensity}
    \nu(x\vert \mu, \Sigma) = \frac{1}{\sqrt{\vert 2\pi \Sigma\vert}} e^{-\frac{1}{2}(x-\mu)^T \Sigma^{-1} (x - \mu)}.
\end{equation}
\end{small}
\vspace{-0.3cm}

The symbol $\nubold$ denotes a set of $n$ Gaussians, 
$\nubold=\{\nu_1,\dots,\nu_n\}$ while $\wbold$ denotes a set of $n$ weights
$\wbold=\{w_1,\dots,w_n\}$, $w_i>0$, so that
the pair $(\wbold,\nubold)$ denotes a Weighted Set of Gaussians (WSG).
The ADF of a Gaussian $\nu$ from a WSG $(\wbold,\nubold)$,
for a given $D$-measure, is defined as in \eqref{eq:mDofp}, and conveniently rewritten as
$m_D(\nu|\wbold,\nubold)$.
The mean and covariance of the Gaussian barycenter are obtained by minimizing the ADF: 

\vspace{-0.1cm}
\begin{small}
\begin{equation} \label{eq:barycenterdef}
\begin{aligned}
(\muhat,\Sigmahat) & = \argmin_{(\mu,\Sigma)\in \Real^d\times S^d}
            m_D(\nu(\cdot |\mu,\Sigma)|\wbold,\nubold)\\
            & = \argmin_{(\mu,\Sigma)\in \Real^d\times S^d}
            \sum_{i=1}^n w_i D\big(\nu_i \Vert \nu(\cdot |\mu,\Sigma)\big),
\end{aligned}
\end{equation}
\end{small}
\vspace{-0.3cm}

\noindent 
and the barycenter is $\nuhat=\nu(\cdot|\muhat,\Sigmahat)$.

\begin{rem} \label{rem:MPM}
A WSG $(\wbold,\nubold)$ can be used to define a Gaussian Mixture (a distribution which is a convex combination Gaussians) as follows:

\vspace{-0.4 cm}
\begin{equation}\label{eq:MoGs}
    p(x|\wbold,\nubold)= \frac{1}{\wbold^T \ones_{n}}
    \sum_{i=1}^n w_i \nu(x\vert \mu_i, \Sigma_i),
\end{equation}
\vspace{-0.2cm}

\noindent
($\wbold^T \ones_n$ is the weight summation).
If a single Gaussian is needed to approximate the mixture, a reasonable choice is
the Gaussian with the same mean and covariance.
In the literature this is called the \emph{Moment-Preserving} merge (MP-merge) of the WSG \cite{Runnalls}.
The mean and covariance of the mixture \eqref{eq:MoGs} are:

\vspace{-0.5cm}
\begin{small}
\begin{align}
    &\muMPM  \triangleq \frac{1}{\wbold^T \ones_{n}}
         \sum_{i=1}^n w_i \mu_i  , \label{eq:muMPM} \\ 
    &\PsiMPM(\muMPM) \triangleq \frac{1}{\wbold^T \ones_{n}}
    \sum_{i=1}^n w_i
               \big(\Sigma_i + (\mu_i - \muMPM)(\mu_i - \muMPM)^T) \label{eq:SigMPM}.
\end{align}
\end{small}
\vspace{-0.2cm}

\noindent
We will see that the Gaussian $\nu(\cdot|\muMPM,\PsiMPM(\muMPM))$ is in fact the barycenter of the
WSG $(\wbold,\nubold)$ according to a specific, well-known, $D$-measure: the Kullback-Leibler Divergence (KLD).
\end{rem}

Usually, the computation of the barycenter of a WSG is not an easy task:
only very few $D$-measures admit a closed form solution for the barycenter, while in most cases the barycenter parameters $(\muhat,\Sigmahat)$ can only be obtained by numerically solving the constrained minimization problem \eqref{eq:barycenterdef}.
In these cases, algorithms like the constrained gradient descent may not be the best choice for two main reasons: 1) the computation of the gradient step while satisfying the constraint on the covariance ($\Sigma\in S^d$) is not an easy task; 2) computationally speaking, addressing the constrained optimization in a gradient descent perspective can be really burdensome, leaving aside all the initialization problems when several local minima are present. 

For these and other reasons, some authors started investigating the use of Fixed Point Iterations (FPI) to achieve fast and robust convergence to the barycenters \cite{FPIJMAA16, ClaiciICML18}, since, if the iterations are well posed, each step will provide a solution in the admitted parameter domain. 
In this paper FPI computations of barycenters are presented for some interesting $D$-measures (see the list in the Appendix),
together with a discussion of their features and peculiarities.

As a general approach, all FPI algorithms presented in this paper are derived by manipulating the
stationary conditions obtained by setting to zero the first order derivatives of the function 
$m_D(\nu|\wbold,\nubold)$ with respect to the parameters of the Gaussian 
$\nu=\nu(x\vert \mu, \Sigma)$.
    In the derivation, the inverse of the covariance,
i.e.,\ the \emph{precision} matrix $\Sigma^{-1}$, has been considered as a parameter,
instead of the covariance $\Sigma$, 
because it provides simpler stationary conditions for all chosen $D$-measures. 
In conclusion, the solution of the following system of equations: 

\vspace{-0.4cm}
\begin{small}
\begin{equation} \label{eq:partialmofnu}
\begin{aligned}
& \frac{\partial m_D(\nu|\wbold,\nubold)}{\partial \mu}  = 
 \frac{1}{\wbold^T\ones_n}
 \sum_{i=1}^n w_i \frac{\partial D\big(\nu_i \Vert \nu(\cdot |\mu,\Sigma)\big)}
{\partial \mu} = 0\\
& \frac{\partial m_D(\nu|\wbold,\nubold)}{\partial \Sigma^{-1}}  = 
 \frac{1}{\wbold^T\ones_n}
 \sum_{i=1}^n w_i \frac{\partial D\big(\nu_i \Vert \nu(\cdot |\mu,\Sigma)\big)}
{\partial \Sigma^{-1}} = 0,\\
& \text{constrained to:}\quad \Sigma\in S^d\quad \big(\Leftrightarrow \quad   \Sigma^{-1}\in S^d\big)
\end{aligned}
\end{equation}
\end{small}
\vspace{-0.2cm}

\noindent is sought for specific choices of $D$-measures.

Throughout the paper, to illustrate the features of the barycenters corresponding to different $D$-measures,
the surface plots of the ADFs over the parameter space $(\mu,\Sigma)$
are presented in Fig. \ref{fig:AllTogether} for the following simple WSG $(\wbold,\nubold)$:

\vspace{-0.0cm} 
\begin{small}
\begin{equation}\label{eq:exampleGaussians}
\begin{gathered}
d=1,\qquad \wbold = \{0.2,0.3,0.4\}, \\
\mubold = \{-1,0,3.5\}, \qquad
\Sigbold = \{0.05,0.1,0.15\}.
\end{gathered}
\end{equation}
\end{small}
\vspace{-0.6cm}

\begin{figure}[hbtp]
    \hspace{-0.3cm}
    \centering\includegraphics[scale=0.15]{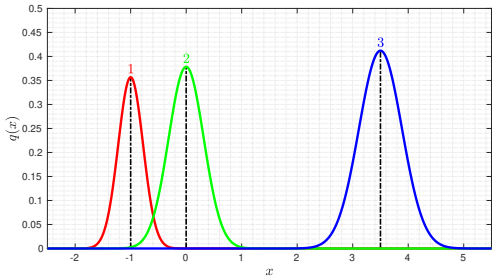}
    \vspace{-0.3 cm}
    \caption{ Plot of the three weighted Gaussians $w_i \nu_i$ of the example \eqref{eq:exampleGaussians}.}
    \label{fig:my_label}
\end{figure}
\vspace{-0.5cm}

\subsection{Forward and Reverse Kullback-Leibler Divergences}

The first $D$-measure to be employed is the \textit{Kullback-Leibler Divergence} \cite{KLD} (KLD) which, given two Gaussian densities $\nu_i=\nu(x\vert \mu_i, \Sigma_i)$ and $\nu_j=\nu(x\vert \mu_j,\Sigma_j)$, is defined as: 

\vspace{-0.3cm}
\begin{small}
\begin{align}
&\DKL(\nu_i\Vert \nu_j) = \int \nu_i(x) \log \frac{\nu_i(x)}{\nu_j(x)}dx  \label{eq:KLDtwoG}
\\ &= \frac{1}{2}\big(\trace(\Sigma_j^{-1} \Sigma_i) + \log \frac{\vert \Sigma_j\vert}{\vert \Sigma_i\vert} 
+(\mu_j - \mu_i)^T \Sigma_j^{-1}(\mu_j - \mu_i) - d \big). \notag
\end{align}
\end{small}

Such divergence satisfies properties \eqref{eq:Dnonneg} and \eqref{eq:Dident}, but it is not symmetric, hence it makes sense to distinguish the \textit{Forward KLD}, defined as in \eqref{eq:KLDtwoG}, from the \textit{Reverse KLD}, obtained by swapping the arguments as $\DRKL(\nu_i\Vert \nu_j) = \DFKL(\nu_j\Vert \nu_i)$. If the $\DFKL$ is employed in \eqref{eq:mDofp}, the system \eqref{eq:partialmofnu} admits a closed form solution that coincides with the \textit{Moment-Preserving} merge described in
Remark \ref{rem:MPM}.
The parameters of the $\DFKL$-barycenter are as follows:

\vspace{-0.2cm}
{\small
\begin{align}
    & \muhat_\FKL =   \frac{1}{\wbold^T \ones_{n}}
    \sum_{i=1}^n w_i \mu_i = \muMPM, \label{eq:FKLDbarymu} \\ 
    & \Sigmahat_\FKL= \frac{1}{\wbold^T \ones_{n}}
    \sum_{i=1}^n w_i
               \big(\Sigma_i + (\mu_i - \muhat_\FKL)(\mu_i - \muhat_\FKL)^T) \label{eq:FKLDbarySigma},
\end{align}
}

\vspace{-1mm}
\noindent
i.e., $\Sigmahat_\FKL=\PsiMPM(\muMPM)$ (see eqs.\ \eqref{eq:muMPM}-\eqref{eq:SigMPM}).
Fig.~\ref{fig:AllTogether}.a reports the plot of the $m_{\DFKL}$ surface as a function of $\mu$ and $\Sigma$ for the simple WSG \eqref{eq:exampleGaussians} and the location of the barycenter parameters.

Computing the $\DRKL$-barycenter, we get what in the literature is known as the \textit{Kullback-Leibler Average (KLA)} \cite{BATTISTELLI2014707} of the Gaussian components:

\vspace{-0.3cm}
\begin{small}
\begin{align}
& \Sigmahat_\RKL= \left(\frac{1}{\wbold^T\ones_{n}}
\sum_{i=1}^n w_i \Sigma_i^{-1} \right)^{-1} \triangleq \PsiKLA,    \label{eq:RKLDbarySigma} \\
& \muhat_\RKL = \Sigmahat_\RKL
\bigg(\frac{1}{\wbold^T\ones_{n}}\sum_{i=1}^n w_i \Sigma_i^{-1} \mu_i\bigg) \triangleq 
            \muKLA(\PsiKLA).
                 \label{eq:RKLDbarymu}
\end{align}
\end{small}

For future purposes, it is also useful to define the following:

\vspace{-0.4cm}
\begin{small}
\begin{equation} \label{eq:mutildeKLA}
\begin{aligned}
\mutildeKLA  \triangleq
\frac{1}{\wbold^T\ones_{n}}\sum_{i=1}^n w_i \Sigma_i^{-1} \mu_i, 
\quad \text{so that}\ \muKLA = \PsiKLA\,  \mutildeKLA. 
\end{aligned}
\end{equation}
\end{small}
\vspace{-0.3cm}

The plot of $m_{\DRKL}$ for the WSG \eqref{eq:exampleGaussians}
and the baricenter parameters are reported in Fig.~\ref{fig:AllTogether}.b.
Note that the $\Sigmahat_{\RKL}$ is significantly smaller than the
variance $\Sigmahat_{\FKL}$ in Fig.~\ref{fig:AllTogether}.a.

To the best of our knowledge, for the Gaussian case these two are the only $D$-measures for which the barycenter parameters can be entirely computed in a closed form. 
For all the other $D$-measures here considered, we provide fixed-point iteration (FPI) algorithms to compute the barycenter parameters.

\subsection{Skew Jeffreys' Divergence}

The Jeffreys' Divergence is a symmetrization of the $KL$-divergence.
Its skew version (see the Appendix) depends on a parameter $\alpha\in[0,1]$, and is the following

\vspace{-0.3cm}
\begin{small}
\begin{equation} \label{eq:DJalpha}
\begin{aligned}
& \DJalpha(\nu_i\Vert \nu_j) =
(1-\alpha)\DFKL(\nu_i\Vert \nu_j) + \alpha\DRKL(\nu_i\Vert \nu_j),
\end{aligned}
\end{equation}
\end{small}

\noindent which for $\alpha=0.5$ recovers the symmetric Jeffreys' Divergence, called also {\it Symmetrized Kullback-Leibler}, denoted $\DSKL$.
Using $\DJalpha$ in \eqref{eq:partialmofnu} two equations are obtained where the unknowns are $(\muhat,\Sigmahat)$ of the $\DJalpha$-barycenter, which do not admit a closed form solution.
After some manipulations, the following iterations are obtained:

\vspace{-0.3cm}
\begin{small}
\begin{equation} \label{eq:DJalphabary}
\begin{aligned}
    & \muhat^{(k+1)} =   (\GammaKLA^{\alpha,(k)})^{-1}\big[\alpha\mutildeKLA  + (\Sigmahat^{(k)})^{-1}(1-\alpha)\muMPM\big], \\
    & \Sigmahat^{(k+1)}= (\GammaKLA^{\alpha,(k)})^{-\frac{1}{2}}\big((\GammaKLA^{\alpha,(k)})^{\frac{1}{2}}
    \GammaMPM^{\alpha,(k)}(\GammaKLA^{\alpha,(k)})^{\frac{1}{2}}\big)^{\frac{1}{2}}(\GammaKLA^{\alpha,(k)})^{-\frac{1}{2}},
\end{aligned}
\end{equation}
\end{small}

\noindent where $\muMPM$ and $\mutildeKLA$ are defined in \eqref{eq:muMPM} and \eqref{eq:mutildeKLA}, and

\begin{small}
\begin{equation}\label{eq:Gammas}
\begin{aligned}
&\GammaKLA^{\alpha,(k)} = \alpha\PsiKLA^{-1} + (1-\alpha)(\Sigmahat^{(k)})^{-1},\\
&\GammaMPM^{\alpha,(k)} = (1-\alpha)\PsiMPM(\muhat^{(k)}) + \alpha \Sigmahat^{(k)}.
\end{aligned}
\end{equation}
\end{small}

\noindent with $\PsiKLA$ defined in \eqref{eq:RKLDbarySigma} and 
$\PsiMPM(\cdot)$ defined in \eqref{eq:SigMPM}. 
Note that, differently from $\PsiKLA$, the matrix $\PsiMPM$ depends on the current mean $\muhat^{(k)}$, and therefore must be recomputed at each iteration. 
If $\alpha=0.5$, the $\DSKL$ case, the equations \eqref{eq:DJalphabary} can be simplified:

\begin{figure*}
\begin{centering}
  \includegraphics[width=0.825\textwidth]{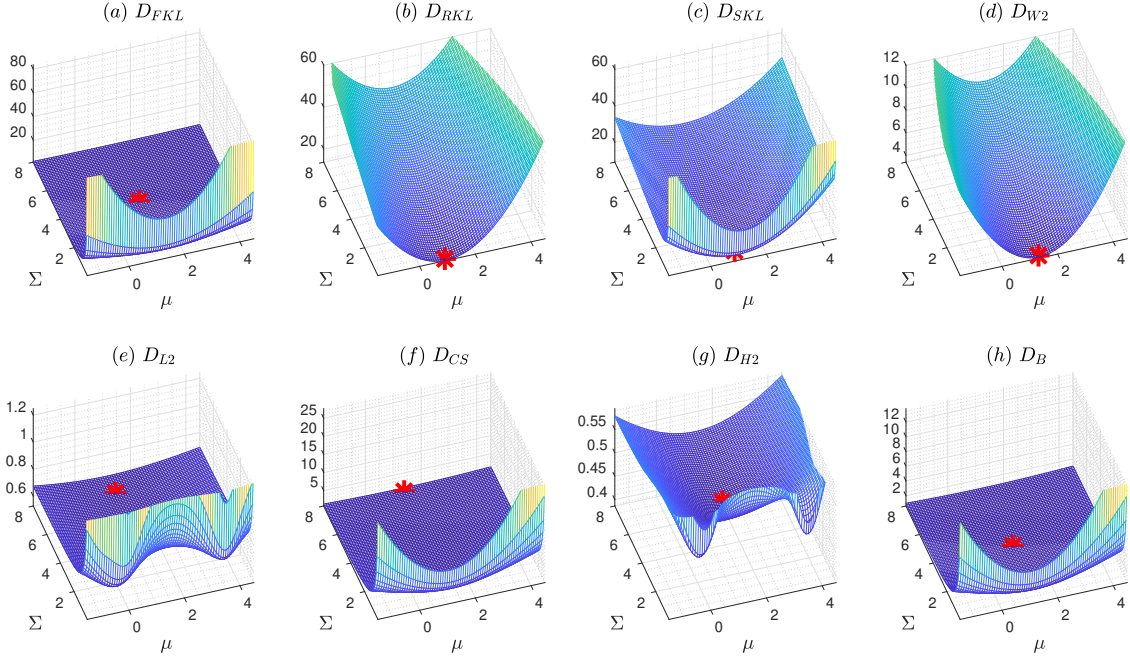}
  \caption{Several $m_D$ plots for the example \eqref{eq:exampleGaussians} over a given grid of the parameters $\mu$ and $\Sigma$. Red star represents the $D$-barycenter values $\muhat$ and $\Sigmahat$ given as (a). $\DFKL$: $\muhat_ {\FKL}=1.3333,\ \Sigmahat_ {\FKL}=4.0000$. (b). $\DRKL$: $\muhat_ {\RKL}=0.5517,\ \Sigmahat_ {\RKL}=0.0931$. (c). $\DSKL$: $\muhat_ {\SKL}=0.6503,\ \Sigmahat_ {\SKL}=0.6449$. (d). $\DW2$: $\muhat_ {W\!2}=1.3333,\ \Sigmahat_{W\!2}=0.1071$. (e). $\DL2$: $\muhat_{L\!2}=1.1571,\ \Sigmahat_{L\!2}=6.8585$. (f). $\DCS$: $\muhat_ {\CS}=1.3239,\ \Sigmahat_ {\CS}=7.7789$. (g). $\DH2$: $\muhat_{H\!2}=1.4968,\ \Sigmahat_{H\!2}=3.7277$. (h). $\DB$: $\muhat_{B}=1.3148,\ \Sigmahat_{B}=3.8914$.}\label{fig:AllTogether}
  \vspace{-0.9cm}
  \end{centering}
\end{figure*}

\vspace{-0.2cm}
\begin{small}
\begin{equation}\label{eq:SKLDbary}
\begin{aligned}
&\muhat^{(k+1)}=\big[\PsiKLA^{-1} + (\Sigmahat^{(k)})^{-1}\big]^{-1}\big[\mutildeKLA  + (\Sigmahat^{(k)})^{-1}\muMPM\big], \\
&\Sigmahat^{(k+1)}= \PsiKLA^{\frac{1}{2}}\big(\PsiKLA^{-\frac{1}{2}}\PsiMPM(\muhat^{(k)})\PsiKLA^{-\frac{1}{2}}\big)^{\frac{1}{2}}\PsiKLA^{\frac{1}{2}},
\end{aligned}
\end{equation}
\end{small}

\noindent which can be efficiently implemented using the Cholesky factorizations of the covariance matrices. 
Note that the above recursions preserve the symmetry and the positive-definiteness of the covariance matrices.
At this stage there is no proof of convergence of the iterations
\eqref{eq:DJalphabary} to the true barycenter mean $\muhat_J$ and covariance $\Sigmahat_J$ 
($\muhat_\SKL$ and $\Sigmahat_\SKL$ for the iterations \eqref{eq:SKLDbary}, for the $\DSKL$ case). 
However, from a broad campaign of tests, and by taking into account the uniqueness of $\DSKL$ barycenters \cite{BregCentroids}, we can say that the proposed FPI algorithm always finds the $\DSKL$-barycenter parameters. 
The iterations can be efficiently initialized by the $\DFKL$-barycenter parameters.
Fig.~\ref{fig:AllTogether}.c reports the $m_{\DSKL}$ surface for the WSG \eqref{eq:exampleGaussians}.

\vspace{-0.0cm}
\subsection{Square 2-Wasserstein Distance}
The Square 2-Wasserstein Distance (W2 for short, see the Appendix) between two Gaussian pdfs $\nu_i=\nu(x|\mu_i,\Sigma_i)$ and $\nu_j=\nu(x|\mu_j,\Sigma_j)$, takes the following closed form
\cite{FPIJMAA16,OTPGMM}:

\vspace{-0.3cm}
\begin{small}
\begin{equation} \label{eq:W2twoG}
\begin{aligned}
& \DW2(\nu_i\Vert \nu_j) =
\Vert \mu_i - \mu_j \Vert_2^2 + tr\big(\Sigma_i + \Sigma_j - 2\big(\Sigma_i^{\frac{1}{2}}\Sigma_j\Sigma_i^{\frac{1}{2}}\big)^{\frac{1}{2}}\big).
\end{aligned}
\end{equation}
\end{small}

By minimizing the average dissimilarity \eqref{eq:barycenterdef} it can be shown
that the parameters of the $\DW2$-barycenter of $(\wbold,\nubold)$ satisfy the following system of equations:

\vspace{-0.3cm}
\begin{small}
\begin{align}
    & \muhat_{W\!2} =   \frac{1}{\wbold^T\ones_{n}}\sum_{i=1}^n w_i \mu_i =\muMPM,  \label{eq:W2barymu} \\
    & \Sigmahat_{W\!2} = \frac{1}{\wbold^T\ones_{n}}
    \sum_{i=1}^n
    w_i\left( \Sigmahat_{W\!2}^{\frac{1}{2}} \Sigma_i 
    \Sigmahat_{W\!2}^{\frac{1}{2}} \right)^\frac{1}{2}. \label{eq:W2barySigma}
\end{align}
\end{small}
\vspace{-0.3cm}

\noindent
Note that \eqref{eq:W2barySigma} is a condition on the covariance that does not allow a closed form solution.
In \cite{FPIJMAA16} the uniqueness of the solution of \eqref{eq:W2barySigma}
has been proven, and the following FPI has been proposed for its computation:

\vspace{-0.2cm}
\begin{small}
\begin{equation}  \label{eq:FPIW2Sigma}
\Sigmahat^{(k+1)}= \frac{1}{\wbold^T\ones_{n}}
    \sum_{i=1}^n
    w_i\bigg(\big({\Sigmahat^{(k)}}\big)^{\frac{1}{2}}\Sigma_i {\big(\Sigmahat^{(k)}\big)}^{\frac{1}{2}}\bigg)^\frac{1}{2}. 
\end{equation}
\end{small}
\vspace{-0.3cm}

Although no proof of convergence has been provided for the FPI \eqref{eq:FPIW2Sigma}
in \cite{FPIJMAA16}, the authors claim its good convergence properties, which we have verified in our extensive numerical tests.
It is interesting to note that a closed form for the $\DW2$-barycenter covariance exists only for $n=2$ \cite{AssaPlataniotis}:

\vspace{-0.3cm}
\begin{small}
\begin{equation*}
    \Sigmahat_{W\!2} = \!\frac{1}{(w_i+w_j)^2}\!\left(w_i^2 \Sigma_i + w_j^2 \Sigma_j
    +w_i w_j\big( (\Sigma_i\Sigma_j)^{\frac{1}{2}}+ (\Sigma_j\Sigma_i)^{\frac{1}{2}}\big)\right).
\end{equation*}
\end{small}
\vspace{-0.3cm}

\noindent 
The plot of the $m_{\DW2}(\mu,\Sigma)$ for the example WSG defined in \eqref{eq:exampleGaussians}
is shown in Fig.~\ref{fig:AllTogether}.d.

For the $D$-measures considered so far, the ADFs $m_D$ appear to suggest the presence of a unique global minimizer (the barycenter); consequently, the proposed FP iterations proposed are quite insensitive to the initialization.
Nonetheless, for several $D$-measures this is not true anymore, and a careful initialization of the FPIs is needed to obtain the convergence to the global minimum of the $m_D$, as happens following.

\subsection{Likeness-based divergence family}

As discussed in \cite{LikenBased}, dissimilarity measures such as the \textit{Square L2 norm}
(aka \textit{Integral Squared Error (ISE)}) and the \textit{Cauchy-Schwarz divergence (CSD)} (see Appendix), belong to the \textit{Likeness-based family}. According to \cite{LikenBased}, a generic \textit{Likeness-based} $D$-measure for two Gaussian pdfs $\nu_i$ and $\nu_j$ can be always put in the form:

\begin{small}
\begin{equation}\label{eq:LBDmeas}
    \DLB(\nu_i\Vert \nu_j)= s_{LB}(J^{i,i},J^{i,j},J^{j,j})
\end{equation}
\end{small}

\noindent 
where $J^{i,i},J^{i,j},J^{j,j}$ are the cross- and self-likeness defined in
\eqref{eq:twoGprodInt} and \eqref{eq:sameGprodInt}, in the Appendix, and $s_{L\!B}$ is a function of those.
When searching for the $\DLB$-barycenter, the following derivatives must be computed

\begin{small}
\begin{equation*}
\begin{aligned}
    &\frac{\partial \DLB(\nu_i\Vert \nu)}{\partial \theta}=
      c_1^{i,\nu}\frac{\partial J^{i,\nu}}{\partial \theta} 
    + c_2^{i,\nu}\frac{\partial J^{\nu,\nu}}{\partial \theta}
\end{aligned}
\end{equation*}
\end{small}

\noindent where the superscript $\nu$ is used in place of $j$ to denote a generic Gaussian density of parameters $(\mu,\Sigma)$, where $\theta\in\{\mu,\Sigma^{-1}\}$, and

\vspace{-0.3cm}
\begin{small}
\begin{equation*}
c_1^{i,\nu}=\frac{\partial s_{LB}(J^{i,i},J^{i,\nu},J^{\nu,\nu})}{\partial J^{i,\nu}},\quad
c_2^{i,\nu}=\frac{\partial s_{LB}(J^{i,i},J^{i,\nu},J^{\nu,\nu})}{\partial J^{\nu,\nu}}.
\end{equation*}
\end{small}
Given a WSG $(\wbold,\nubold)$ and a Gaussian $\nu$ let us define:

\vspace{-0.3cm}
\begin{small}
\begin{equation*} 
\begin{aligned}
\wtilde_{i,\nu} & = w_i J^{i,\nu}\ , &  \wtilde^{c_1}_{i,\nu} &=  w_i J^{i,\nu} c_1^{i,\nu}, &
\wtilde^{c_2}_{i,\nu} &= w_i J^{\nu,\nu}c_2^{i,\nu},\\
\wbar_\nu & = \sum_{i=1}^n\wtilde_{i,\nu}\ , &
\wbar^{c_1}_\nu & = \sum_{i=1}^n\wtilde_{i,\nu}^{c_1}\ ,
& \wbar^{c_2}_\nu & = \sum_{i=1}^n\wtilde_{i,\nu}^{c_2}\ .
\end{aligned}
\end{equation*}
\end{small}

After some manipulation of the stationary equations, the following general FPI algorithm for the barycenter computation in the Likeness-based family have been derived:

\vspace{-0.3cm}
\begin{small}
\begin{align}
\hspace{-0.2cm}
&\muhat^{(k+1)} = \frac{1}{\wbar_{\nu}^{c_1}} \sum_{i=1}^n \wtilde_{i,\nu}^{c_1} \mubar_{i,\nu} , \label{eq:LBbarycenter}\\
&\Sigmahat^{(k+1)} = \frac{1}{\wbar_{\nu}^{c_1}}\Big[\sum_{i=1}^n \wtilde_{i,\nu}^{c_1}\big(\Sigmabar_{i,\nu} + (\mubar_{i,\nu}-\muhat^{(k)})(\mubar_{i,\nu}-\muhat^{(k)})^T\big) \notag\\
& \hspace{3cm} - \wbar_{\nu}^{c_2}\Sigmahat^{(k)}\Big], \notag
\end{align}
\end{small}
\vspace{-0.3cm}

\noindent where $\nu$ in the subscripts denotes the Gaussian barycenter varying at each iteration, hence function of $(\muhat^{(k)},\Sigmahat^{(k)})$:
for simplicity of notation the dependency of $\nu$ on $k$ has been omitted. 
$\mubar_{i,\nu}$ and $\Sigmabar_{i,\nu}$ are computed as in \eqref{eq:muSigmabar} (see Appendix).
For the $\DL2$-measure, we have $c_1^{i,\nu}=-2$ and $c_2^{i,\nu}=1$ \cite{LikenBased}, and  
the above recursion can be rewritten as:

\vspace{-0.3cm}
\begin{small}
\begin{align}
&\muhat^{(k+1)} = \frac{1}{\wbar_{\nu}} \sum_{i=1}^n \wtilde_{i,\nu} \mubar_{i,\nu}, \label{eq:L2barycenter} \\
&\Sigmahat^{(k+1)} = \frac{1}{\wbar_{\nu}}\Big[\sum_{i=1}^n \wtilde_{i,\nu} \big(\Sigmabar_{i,\nu} + (\mubar_{i,\nu}-\muhat^{(k)})(\mubar_{i,\nu}-\muhat^{(k)})^T\big) \notag\\
& \hspace{3cm} + \frac{1}{2}\wbold^T\ones_n J^{\nu,\nu}\Sigmahat^{(k)}\Big].
\notag
\end{align}
\end{small}

\vspace{-0.1cm} \noindent
Fig.~\ref{fig:AllTogether}.e reports the plot of $m_{\DL2}$ and the barycenter parameters
$\muhat_{L\!2}$ and $\Sigmahat_{L\!2}$ for the WSG example \eqref{eq:exampleGaussians}. 
Note that the $m_{\DL2}$ functions presents local minima, and that the global minimum is quite far from them.

Another $D$-measure in the LB family we here consider is the \textit{Cauchy-Schwarz Divergence}
$D_\CS$,
for which $c_1^{i,\nu} = -\frac{1}{J^{i,\nu}}$ and $c_2^{i,\nu} = \frac{1}{2J^{\nu,\nu}}$
(see \cite{LikenBased}).
By exploiting the general form \eqref{eq:LBbarycenter}, we get the following recursion:

\vspace{-0.3cm}
\begin{small}
\begin{align}
&\muhat^{(k+1)} = \frac{1}{\wbold^T\ones_n}\sum_{i=1}^{n}w_i \mubar_{i,\nu}, \label{eq:CSDbarycenter}\\
&\Sigmahat^{(k+1)} = \frac{2}{\wbold^T\ones_n}\sum_{i=1}^n w_i\big(\Sigmabar_{i,\nu} + (\mubar_{i,\nu}-\muhat^{(k)})(\mubar_{i,\nu}-\muhat^{(k)})^T\big). \notag
\end{align}
\end{small}

Fig.~\ref{fig:AllTogether}.f reports the plot of $m_{\DCS}$ and the barycenter parameters
$\muhat_\CS$ and $\Sigmahat_\CS$.

\subsection{Chernoff $\alpha$-divergences}
The Chernoff $\alpha$-divergences have been defined in 
\eqref{eq:DalphadivI}-\eqref{eq:DalphadivII} in the Appendix,
for $\alpha\in (-\infty,\infty)$, and it is known that for some values of $\alpha$
they coincide with some of the listed divergences.
Here we limit ourselves to the case $\alpha \in [0,1]$, since for $\alpha \not\in [0,1]$ 
the divergence is not well defined for all pairs of Gaussians.\\
The {\it $\alpha$-divergence of the $I^\circ$ kind} for a pair $\nu_i$, $\nu_j$
is 

\begin{small}
\begin{equation}\label{eq:Dalpha}
\DaP(\nu_i\Vert \nu_j) = \frac{1}{\alpha (1-\alpha)} \big( 1 - c_{\alpha}(\nu_i,\nu_j)\big),
\end{equation}
\end{small}

\noindent where $c_{\alpha}(\nu_i,\nu_j)$, $c_{\alpha}^{i,j}$ for short, is the Chernoff coefficient defined in \eqref{eq:Chernoff},
which for Gaussians takes the form \eqref{eq:geomTwoG}.
As done with the LB family, let us define the quantities:

\vspace{-0.2cm}
\begin{small}
\begin{equation}\label{eq:Calphaweights}
w^{c_\alpha}_{i,\nu} = w_i c_{\alpha}^{i,\nu},\qquad 
\wbar^{c_\alpha}_{\nu}=\sum_{i=1}^n w^{c_\alpha}_{i,\nu}.
\end{equation}
\end{small}

By using $\DaP$ in \eqref{eq:mDofp}, we have been able to formulate the following FPI:

\vspace{-0.3cm}
\begin{small}
\begin{align}
&\muhat^{(k+1)} = \frac{1}{\wbar^{c_\alpha}_{\nu}}
                \sum_{i=1}^{n} w^{c_\alpha}_{i,\nu} \mubar_{i,\nu}^{\alpha}, \label{eq:DaPbarycenter}\\
&\Sigmahat^{(k+1)} = \frac{1}{\wbar^{c_\alpha}_{\nu}}
                \sum_{i=1}^n w^{c_\alpha}_{i,\nu} \big(\Sigmabar_{i,\nu}^{\alpha} + (\mubar_{i,\nu}^{\alpha}-\muhat^{(k)})(\mubar_{i,\nu}^{\alpha}-\muhat^{(k)})^T\big), \notag
\end{align}
\end{small}
\vspace{-0.3cm}

\noindent  
where $\mubar_{i,\nu}^{\alpha}$ and $\Sigmabar_{i,\nu}^{\alpha}$ are
defined in \eqref{eq:muSigmabarAlpha} in the Appendix; again, the dependence of $\nu$ on $k$ has been omitted.
In Fig.~\ref{fig:AllTogether}.g the $m_D$ surfance for $\alpha=\frac{1}{2}$
(\textit{Square Hellinger Distance}) for the example 
\eqref{eq:exampleGaussians}, is reported.

Like for the Square L2 norm, also for this class of $\alpha$-divergences the function
$m_{\DaP}$ may present local minima when $\alpha\in(0,1)$. 
Hence, also in this case it is important to carefully select the initial guess for the FPI. 

The {\it $\alpha$-divergences of the $II^\circ$ kind}, defined in \eqref{eq:DalphadivII} in the Appendix, when applied to pairs of Gaussian is:

\vspace{-0.1cm}
\begin{small}
\begin{equation}\label{eq:skewBhatt}
    \DaS(\nu_i\Vert \nu_j) = -\log c_{\alpha}(\nu_i,\nu_j),
\end{equation}
\end{small}
\vspace{-0.4cm}

\noindent with $c_{\alpha}(\nu_i,\nu_j)$ taking the form \eqref{eq:geomTwoG}.
As reported in the Appendix, for $\alpha=0.5$, $\DaS$ coincides with the \textit{Bhattacharyya} distance. 
One nice property of this class of $\alpha$-divergences, is that the fixed-point iteration for the barycenter computation results to be simpler if compared to the divergences of the first kind, mainly because the Chernoff coefficient is cancelled out.
The fixed-point iteration derived for the $\DaS$-barycenter is:

\vspace{-0.5cm}
\begin{small}
\begin{equation}
\begin{aligned}
&\muhat^{(k+1)} = \frac{1}{\wbold^T\ones_n} \sum_{i=1}^{n} w_i \mubar_{i,\nu}^{\alpha},\label{eq:DaSbarycenter}\\
&\Sigmahat^{(k+1)} = \frac{1}{\wbold^T\ones_n} \sum_{i=1}^n w_i\big(\Sigmabar_{i,\nu}^{\alpha} + (\mubar_{i,\nu}^{\alpha}-\muhat^{(k)})(\mubar_{i,\nu}^{\alpha}-\muhat^{(k)})^T\big).
\end{aligned}
\end{equation}
\end{small}
\vspace{-0.2cm}

In Fig.~\ref{fig:AllTogether}.h is reported the $m_{\DaS}$ surface for $\alpha=0.5$, which corresponds to 
$m_{\DB}$, where $\DB$ is the \textit{Bhattacharyya} distance.

\section{Discussion}\label{sec:discussion}

\hide{In the previous section several fixed-point iterations (FPI) have been reported for different $D$-measures, but little has been said regarding the initialization, the stopping criteria and convergence. The example \eqref{eq:exampleGaussians} we have reported is particularly simple, thus it should not be taken as a way to figure out where and when the algorithm will converge, given that the parameter space gets more and more complicated while the dimensionality of the problem increases. The only way to figure out if the barycenter solution is unique for a given $D$-measure should be by investigating analytically the general case, even if it might be really difficult from several points of view. If we consider the $\alpha$-divergences of the second kind, they belong to the Burbea-Rao class \cite{BurbeaRao}, for which it is proven that the barycenter exists and it is unique. Thus, we expect the algorithm to always converge; still, the initialization might influence the convergence time. One thing that we empirically observed is that the $\DB$ has similar features to the $\DFKL$, hence it may be reasonable to choose as initial guess for the corresponding FPI the parameters obtained as in \eqref{eq:FKLDbarymu} and \eqref{eq:FKLDbarySigma}.
} 

When considering Fixed-Point Iterations (FPI) like those presented in the previous section, important issues to be considered are the algorithm initialization, the convergence properties, the possibility of multiple stationary points, and the stopping criteria. 
All FPIs presented have been constructed so that they are stationary when the parameters $(\mu,\Sigma)$ satisfy the stationary conditions \eqref{eq:partialmofnu}
for the Average Dissimilarity Function (ADF).
It must be noted that the existence of more than one stationary point for the ADF, and in particular of local minima, depends on the $D$-measure considered. 
However, in most cases it is not easy to analytically investigate the presence of local minima.
For instance, considering the $\alpha$-divergences of the second kind, like the Bhattacharyya distance $D_B$,
we know that they belong to the Burbea-Rao class  for which the barycenter exists unique, as proven in \cite{BurbeaRao}.
Although this property does not exclude the presence of local minima for the ADF, it suggests that a proper initialization of the FPI algorithm can ensure the convergence to the barycenter.
Considering the Bhattacharyya distance $D_B$, the numerical tests we have performed have shown that its features are rather similar to the $D_\FKL$, hence it is reasonable to choose as initial guess for the
FPI the parameters of $D_\FKL$-barycenter, i.e., the \textit{MP}-merge 
\eqref{eq:muMPM}-\eqref{eq:SigMPM}.
Indeed, in our campaign of numerical tests we have found that under this initialization the FPI algorithm  for the $D_B$-barycenter always converges.

One interesting link between the Bhattacharyya and the Cauchy-Schwarz FPI iterations, 
\eqref{eq:CSDbarycenter} and \eqref{eq:DaSbarycenter}, with $\alpha=0.5$,
is that the covariance iterations differ by a factor 2 
(the uniqueness for the $\DCS$-barycenters is very likely, although, to the best of our knowledge, there is no proof of that at this moment). 
Thus, for the FPI \eqref{eq:CSDbarycenter} for the $\DCS$-barycenter we considered the same inizialization \eqref{eq:CSDbarycenter} and \eqref{eq:DaSbarycenter} used for the 
$\DB$-barycenter FPI.

The FPI \eqref{eq:FPIW2Sigma} for the $\DW2$-barycenter has been proposed in \cite{OTPGMM}, without any proof of convergence, with the support of successful numerical tests. 
Regarding the initialization, since the $\DW2$ measure is somehow similar in the features to the $\DRKL$, a reasonable initial guess would be as in \eqref{eq:RKLDbarymu} and \eqref{eq:RKLDbarySigma}.

As discussed in \cite{BregCentroids}, the $\DSKL$ barycenters are unique. 
Also in this case, in our numerical test we used the moment-preserving merge parameters (i.e., the $\DFKL$-barycenter) as starting point, always obtaining fast convergence.

Regarding the $\DaP$-barycenters, we mentioned that for $\alpha \in (0,1)$ the solution might not be unique. From several tests, there has not been a prevalent heuristic in the choice of the initial guess. 
However, we observed that measures as the $\DH2$ favor the preservation of the main distribution peaks, hence, at first, we found reasonable to consider as starting point the parameters of the component with the highest peak (largest ratio $w_i/|\Sigma_i|$). 
Nonetheless, close to the main peaks may be local minima, and by initializing the recursion on those, the algorithm often sticks with the corresponding, potentially sub-optimal, solution. As an alternative, we used again the moment-preserving merge parameters as a starting point
($\DFKL$-barycenter), which in general is an inclusive solution, thus leaving more chance to converge to a different minimum w.r.t.\ the ones associated with the main peaks. An alternative could be to evaluate the ADF either on the solution provided by an \textit{MP}-merge initialization or by the one initialized on the main peak; the solution yielding the least value can be accepted as $\DaP$-barycenter. 

The same argumentation can be extended to measures as the Square L2 norm, for which there are several local minima when seeking the barycenter problem solution.

To conclude this section, the stopping criteria are shortly discussed. 
Given that, in general, fixed-point iteration algorithms asymptotically converge to a solution, one should choose a maximum number of allowed iterations and a tolerance on the approximation accuracy. Regarding the former, we have experimented that some algorithms, in particular the FPIs for the  $\DH2$ and $\DL2$ barycenters, may take many iterations to converge when dealing with high-dimensional problems; for this reason, one might think to either permit a large number of iterations, or allow for smaller accuracy in the resulting approximation. Regarding the tolerance, considering that, in the Gaussian case, we are looking for the mean and the covariance parameters, one could consider to check the accuracy by evaluating the variation in norm between the parameters updates. Otherwise, one could consider the divergence between the barycenter at the previous iteration, and the new one obtained from a recursion step. In both cases, it is important to select the tolerance order of magnitude accordingly.

\vspace{-0.6cm}
\section{Conclusions}
In this work Fixed-Point Iteration (FPI) algorithms are provided for the computation of barycenters of Weighted Sets of Gaussians (WSG) according to several dissimilarities ($D$-measures).
The barycenter has been defined as the minimizer of an {\it average dissimilarity function} (ADF).
Through some mathematical manipulations, it has been possible to obtain compact formulations of FPIs that solve the minimization problem for a wide set of $D$-measures. 
It has been stressed that, depending on the chosen $D$-measure, the ADF may have local minima or not.
To help visualize such cases, the plots of the ADFs of the considered $D$-measures have been reported for a simple WSG chosen as an example. 
Initialization and convergence have been discussed for each $D$-measure.
As a future work, we are planning to try to provide theoretical convergence guarantees for the proposed FPIs.

\vspace{-0.3cm}
\section{Appendix}

The list of divergences ($D$-measures) used in this paper is reported below
(all ratios between pdfs are assumed well defined):

\vspace{-0.3cm}
\begin{small}
\begin{align*}
\label{eq:someDmeasA}
& \text{Kullback-Leibler} & \DKL(p\Vert q)  =\int p(x)\log \frac{p(x)}{q(x)}\d x,\\ 
& \text{Jeffreys (symm-KL)} & 
        \DJ(p\Vert q) = \frac{1}{2}(\DKL(p\Vert q) + \DKL(q\Vert p))\\ 
& &          = \frac{1}{2}\int\! (p(x)\! -\! q(x))\! \log \frac{p(x)}{q(x)} \d x,\\ 
& \underset{\text{$\alpha\in[0,1]$}}{\text{Skew Jeffreys}} & 
        \DJ^\alpha(p\Vert q) =  (1-\alpha)\DKL(p\Vert q) +\alpha \DKL(q\Vert p), \\
& \underset{\text{aka Integral Square Error}}{\text{Square L2 norm}} & 
    \DL2(p\Vert q)  = \int \big(p(x)-q(x)\big)^2 \d x,\\
& \text{Cauchy-Schwarz} &
\DCS(p||q) = - \log \frac{\int p(x)q(x) \d x}{\sqrt{\int p^2(x)\d x \int q^2(x)\d x}}, \\
& \text{Bhattacharyya} & 
    D_B(p\Vert q) = - \log \int\! \sqrt{p(x)q(x)}\d x, \\
& \text{Square Hellinger} & 
    \DH2(p\Vert q) =\frac{1}{2} \int\! \left(\!\sqrt{p(x)}\!-\!\sqrt{q(x)}\right)^2\!\d x\\
& & = 1 - \int \sqrt{p(x)q(x)}\d x,\\
& \text{Sq. 2-Wasserstein} & \DW2(p\Vert q)=\inf_{\pi\in\mathcal{Q}_2} \int\!\!\int \Vert x-y\Vert^2 \pi(x,y)\d x \d y
\end{align*}
\end{small}

\noindent where $\mathcal{Q}_2$ is the set of all pdfs in $\Real^d\times\Real^d$ that have
$p(x)$ and $q(y)$ as marginals and finite second order moments.

\medskip

\noindent
{\it $\alpha$-divergences}\\[3pt]
The Chernoff $\alpha$-coefficient $c_\alpha(p,q)$, $\alpha\in (-\infty,\infty)$, defined as

\vspace{-0.5cm}
\begin{small}
\begin{equation} \label{eq:Chernoff}
c_\alpha(p,q)=\int p^\alpha(x) q^{1-\alpha}(x)\d x
\end{equation}
\end{small}
\vspace{-0.4cm}

\noindent allows to define two families of divergences:
$\alpha$-divergences of the $I^\circ$ and of then $I\!I^\circ$ kind:

\vspace{-0.3cm}
\begin{small}
\begin{align}
& \text{$I^\circ$ kind $\alpha$-div} &
\DaP(p\Vert q) & = \frac{1}{\alpha (1-\alpha)} \big( 1 - c_{\alpha}(p,q)\big), \label{eq:DalphadivI} \\
& \text{$I\!I^\circ$ kind $\alpha$-div} &
\DaS(p\Vert q) & = - \log c_{\alpha}(p,q). \label{eq:DalphadivII}
\end{align}
\end{small}

For some values of $\alpha$, the  $\alpha$-divergences of the $I^\circ$ kind coincide with some of the previously listed divergences:

\begin{small}
\begin{equation}
\begin{aligned}
\DaP(p\Vert q)\big\vert_{\alpha=-1}& =\DP(p\Vert q) & \text{Pearson $\chi^2$}\\
\lim_{\alpha\to 0} \DaP(p\Vert q) & =\DKL(q\Vert p) & \text{Reverse KL}\\
\DaP(p\Vert q)\big\vert_{\alpha=0.5}& =4\DH2(p\Vert q) & \text{Square Hellinger}\\
\lim_{\alpha\to 1} \DaP(p\Vert q) & =\DKL(p\Vert q) & \text{Forward KL}\\
\DaP(p\Vert q)\big\vert_{\alpha=2}& =\DN(p\Vert q) &  \text{Neyman $\chi^2$}
\end{aligned}  
\end{equation}
\end{small}

For $\alpha=0.5$, $c_\alpha(p,q)$ coincides with the Bhattacharayya coefficient \cite{BurbeaRao}
and the corresponding $\alpha$-divergence of the $I\!I^\circ$ kind coincides with the
Bhattacharayya distance.

\vspace{-0.1cm}
\subsection{Some useful formulas}\label{sec:someIdentities}

The \textit{cross-likeness} $J^{i,j}$ between two Gaussians $\nu_i=\nu(x\vert \mu_i,\Sigma_i)$ and $\nu_j=\nu(x\vert \mu_j,\Sigma_j)$ is defined in \cite{LikenBased} as

\vspace{-0.1cm}
\begin{small}
\begin{equation} \label{eq:twoGprodInt}
J^{i,j}=\int \nu_i(x) \cdot \nu_j(x) dx = \nu(\mu_i\vert \mu_j, \Sigma_i + \Sigma_j).
\end{equation}
\end{small}
\vspace{-0.4cm}

The \textit{self-likeness} of a Gaussian $\nu_i$ is

\vspace{-0.1cm}
\begin{small}
\begin{equation}\label{eq:sameGprodInt} 
J^{i,i}=\int\! \nu^2(x) dx = \nu(\mu_i\vert \mu_i,2\Sigma_i)=\vert 4\pi\Sigma_i\vert^{-\frac{1}{2}}.
\end{equation}
\end{small}
\vspace{-0.4cm}

If $\nu$ is without subscript, then the self-likeness is written $J^{\nu,\nu}$, so that 
$J^{\nu,\nu}=\nu(\mu\vert \mu, 2\Sigma)$, and the cross-likeness between $\nu_i$ and $\nu$
is written as $J^{i,\nu}$. 
The following hold true
\begin{align}
& \nu_i(x) \cdot \nu(x) = J^{i,\nu}\,\nu(x\vert \mubar_{i,\nu}, \Sigmabar_{i,\nu}), 
\label{eq:twoGprod}\\
&\text{where}\qquad
\begin{aligned}
& \mubar_{i,\nu} = \big(\Sigma_i^{-1} + \Sigma^{-1}\big)^{-1}\big(\Sigma_i^{-1}\mu_i + \Sigma^{-1}\mu\big) ,\\
& \Sigmabar_{i,\nu} = \big(\Sigma_i^{-1} + \Sigma^{-1}\big)^{-1},
\end{aligned}
\label{eq:muSigmabar}
\end{align}

\vspace{-0.2cm}
\begin{small}
\begin{equation}\label{eq:sameGprod}
    \nu^2(x) = J^{\nu,\nu}\,\nu\left(x\big\vert \mu, \textstyle{\frac{1}{2}}\Sigma\right)=\frac{1}{\sqrt{\vert 4\pi\Sigma\vert}}\nu\left(x\big\vert \mu, \textstyle{\frac{1}{2}}\Sigma\right).
\end{equation}
\end{small}
\vspace{-0.2cm}

For any $\alpha \in (0,1)$ we have:

\begin{small}
\begin{equation}\label{eq:alphaPowG}
\begin{aligned}
& \nu^{\alpha} = \frac{(2\pi)^{\frac{d}{2}(1-\alpha)} \vert\Sigma\vert^{\frac{1-\alpha}{2}}}{\alpha^{\frac{d}{2}}}
        \nu\left(x\vert \mu, \textstyle{\frac{1}{\alpha}}\Sigma\right),\\
& \nu^{1-\alpha} = \frac{(2\pi)^{\frac{d}{2}\alpha} \vert\Sigma\vert^{\frac{\alpha}{2}}}{(1-\alpha)^{\frac{d}{2}}}\nu
            \left(x\big\vert \mu, \textstyle{\frac{1}{1-\alpha}}\Sigma\right).
\end{aligned}
\end{equation}
\end{small}

The Chernoff $\alpha$-coefficient \eqref{eq:Chernoff}
between $\nu_i$ and $\nu_j$ is

\vspace{-0.3cm}
\begin{small}
\begin{equation} \label{eq:alphaCtwoG} 
c_{\alpha}(\nu_i,\nu_j)=
\frac{\vert \Sigmabar_{i,j}^{\alpha}\vert^{\frac{1}{2}}}{\vert \Sigma_i\vert^{\frac{\alpha}{2}} \vert\Sigma_j\vert^{\frac{1-\alpha}{2}}} e^{-\frac{1}{2}(\mu_i-\mu_j)^T(\Sigtilde_{i,j}^{\alpha})^{-1}(\mu_i-\mu_j)},
\end{equation}
\end{small}
\vspace{-0.3cm}

\vspace{-0.4cm}
\begin{small}
\begin{align}
\text{so that}\quad & \nu_i^{\alpha}\cdot \nu^{1-\alpha}
    =c_{\alpha}(\nu_i, \nu)\cdot \nu(x\vert \mubar_{i,\nu}^{\alpha},\Sigmabar_{i,\nu}^{\alpha}),
    \label{eq:geomTwoG}\\
& \text{where}\qquad
\begin{aligned}
& \Sigmabar_{i,\nu}^{\alpha} = \big(\alpha\Sigma_i^{-1} + (1-\alpha)\Sigma^{-1}\big)^{-1}, \\
& \Sigtilde_{i,\nu}^{\alpha} = 
\textstyle{\frac{1}{\alpha}}\Sigma_i + \textstyle{\frac{1}{1-\alpha}}\Sigma,\\
& \mubar_{i,\nu}^{\alpha} = \Sigmabar_{i,\nu}^{\alpha}\,\big(\alpha\Sigma_i^{-1}\mu_i + (1-\alpha)\Sigma^{-1}\mu\big).
\end{aligned}
\label{eq:muSigmabarAlpha}
\end{align}
\end{small}



\bibliographystyle{IEEEtran}
\bibliography{IEEEabrv,main}

\end{document}